# Blockchain-Based Electronic Voting System for Elections in Turkey

Rumeysa Bulut, Alperen Kantarcı, Safa Keskin, Şerif Bahtiyar
*Faculty of Computer and Informatics*
*Istanbul Technical University*
Istanbul, Turkey
{bulutr, kantarcia, keskinsaf, bahtiyars}@itu.edu.tr

*Abstract*—Traditional elections satisfy neither citizens nor political authorities in recent years. They are not fully secure since it is easy to attack votes. It threatens also privacy and transparency of voters. Additionally, it takes too much time to count the votes. This paper proposes a solution using Blockchain to eliminate all disadvantages of conventional elections. Security and data integrity of votes is absolutely provided theoretically. Voter privacy is another requirement that is ensured in the system. Lastly, waiting time for results decreased significantly in proposed Blockchain voting system.

*Keywords—blockchain, e-voting, electronic voting, internet voting*

## I. INTRODUCTION

Elections are the organizations that is supposed to bring democracy into countries. They mostly play a crucial role in the future of a country and citizens life. Therefore, it has much importance for every single person involved in these elections. Regardless of the organization, elections have to be trustworthy in its nature. They have to ensure people's privacy and vote's security. Additionally, the authority which is responsible for counting votes should not spend too much time on counting votes since waiting long period of time for results increases concerns about manipulation of results. However, due to the different reasons depending on the areas that elections have been made, trust has been a controversial issue for each election. Especially, paper elections are managed by a centralized authority, there is always a risk to manipulate ballots and election results [1]. For example, there are disagreements about security and privacy of elections and whether all votes are counted correctly in Turkey. Moreover, time factor can be a challenging issue for announcing results. The speed of counting votes and making public the unofficial results takes seven to eight hours in Turkey, whereas the official results are announced in ten days approximately. The official results of 24th June General Elections of Turkey were expounded after eleven days on 5th July.

There are some attempts to remove problems of traditional election system. These attempts try to benefit from online systems to automate the whole process. Electronic voting was used in elections of Austrian Federation of Students in 2009 [2] and in some elections in Switzerland [3]. Although e-voting makes selection operation easy, privacy and security worries still continue. To dissipate problems of both conventional and e-voting elections, e-voting can be improved using Blockchain mechanism. Blockchain has impressive features to overcome troubles of voter's security, privacy and data integrity of votes.

Blockchain is an inalterable and an easy confirmable system [4]. Under favor of these qualifications, Blockchain has a significant potential to be an alternative to traditional elections. It brings smart solutions to central authority problem in terms of all blocks having all data in the chain. Also, it is impossible to change an information in a block since it is discerned by other blocks which have whole data. Consequently, Blockchain increases the security of information by keeping the entire data in all blocks, and removes the need for an official center to provide a secure election [5]. As mentioned before, counting votes and making election results publicly available takes considerable time. Blockchain solves this problem by its nature. Since the last node on the chain keeps all information, it is enough to look for only the last node for the results. This reduces the waiting time dramatically. Thus, incomplete and official results are explained at the same time.

Since each country has different laws and implementations, proposing a definitive structure is almost impossible. The suggested solution in this paper is specifically for problems of conventional paper elections in Turkey. Despite the solution is specific to one country, it may be taken as a general application, and can be customized to other countries. Security and privacy of votes and voters and the speed of counting votes and announcing the results are discussed in the solution. A representative model of the system is presented in Proposed Solution section.

The paper is organized as follows. In Section II, related works about e-voting and blockchain are discussed. In Section III, general architecture of the proposed blockchain based e-voting election system is explained and modeled with supporting materials. At the following section, Section 4, proposed system is analyzed from different aspects. In the final section, Section 5, future work on the system is discussed and conclusions about the research have been made.

## II. E-VOTING SYSTEMS AND BLOCKCHAIN

E-voting in different areas has been considered for a while by some countries. The pioneer country in e-voting process is Estonia which held online voting between 2005 and 2007 [6]. On the other hand, Blockchain-based election has not been commonly applied yet. It is in development process in recent years. South Korea is a noticeable specimen which brought Blockchain-based election to a successful conclusion in 2017 [7]. Additionally, there are some papers that widen viewpoint about Blockchain-based voting by offering different ideas.

However, not all of them is found useful by authors of this paper.

In [8], the voting process relies on citizen's email address that can be hacked or manipulated easily. To be obvious, there will be always some people who registers to the system using someone else's mail address and votes on behalf of them. For example, a grandson may open an email address for his grandparents from different devices, and cast their votes. This method guarantees none of the required qualifications such as security, data integrity or privacy that an e-voting system has system. For such a system, stealing votes or changing votes are totally.

Researchers of the [9], proposed a peer-to-peer blockchain based voting system. Main focus of this research is to protect the anonymity of the ballots and commitment of the votes to the blockchain. According to this purpose, they propose a unique vote commitment format. Their solution has solid base for such a vote commitment format but we propose a different system that leans on another system that is maintained by government. In this purpose we preferred to use a structure for chains that consists of different key-value pairs represents vote itself.

Another paper proposes a solution with a database alongside Blockchain [1]. The authors designed a system that creates blocks following collection of ballots from voters to keep them in a database until the end of election process [1]. In this paper, authors tried to eliminate the need for a database.

Blockchain is being used for various areas such as IoT. Since there are so many different devices and each of them are processing different data, new approaches emerged from this area. In [10] author discusses the blockchain IoT interactions in the paper and one of the model that have been discussed is hybrid model that uses different chains in different layers and levels which inspired us in our blockchain based e-voting system.

Another work in the related area propose a one-time ring signature in order to ensure the anonymity of the voting citizen [11]. However, each candidate in the election needs public key pair in this architecture and adding a new candidate increase the complexity of signing process and at every node demanded CPU power increases. One time ring signature architecture does not depend any trusting center but in our special case we give the authority of the selecting a candidate to the government which is the trusting center for the election.

Previous researches are reported according to the most remarkable properties in Table 1.

### III. BLOCKCHAIN-BASED VOTING

Considering today's technology, blockchain may create one of the most prominent alternative to traditional voting in terms of security, consistency and speed. While designing a chain for voting in a crowded country, the system should be secure. Many aspects should be considered in order to construct a secure blockchain-based election system. First factor is human for such a system. In the solution, human interference is absolutely prohibited.

The proposed system will be consisting of nodes (computers in design) that is closed to human interference. Any input that cannot be considered as vote will be ignored in this system. For such a system, stealing votes or changing votes are totally blocked. Second issue is saving system from hackers. In order to manipulate votes, hackers need to enter the system as a citizen at proposed solution. Also, it is guaranteed that a citizen can only vote for one time. When citizen cast a ballot, e-government system will be informed without revealing any information about vote. Then, e-government system marks that person as voted. Since the system takes electorate data from e-government, it is not possible for a marked person to vote again. Although a hacker is obtained the citizen information and entered to the system, he cannot vote more than one time.

In a blockchain system, every transaction is related to the previous one. So, changing an accepted transaction is impossible for such a system. Due to the consistency of the blockchain, data will always be consistent and voting will be reliable. In a case of manipulation of the system such as changing votes or stealing votes, other connected nodes will already be synchronized. So, the changed data will be identified instantly. Details of the system will be explained below after the use case diagram and explanation of it.

As you can see in the Fig. 1, standard use case of the election system is about the citizen and government. Government in this system only provides the authorization of the citizens who can vote or prevention of the citizens who already voted for that election. Also, government and citizens determine the candidates that will be participating in that election. The ballot box information, candidates and citizen ballot box relation will be provided by the government which is the trusted party in the elections. After citizen's vote, it is added to the blockchain that we will be proposed below and any vote has a guarantee from the system about being immutable. Since a chain contains all the citizen votes anonymously at the end of the election, the official results will be announced within minutes after the election terminates. Any concerning third party can get the chain and count the votes for being sure that voting is really trusted.

TABLE I: METHODS OF PREVIOUS WORKS

| Researches | Architecture and Design | Security Considerations | User Authentication |
|---|---|---|---|
| [1] | • Blockchain permission<br>• Maintain data integrity | • Data integrity<br>• Digital Signature | • Asymmetric authentication |
| [4] | • Smart contract<br>• POA Permissioned blockchain<br>• Exonum, Quorum and Geth Frameworks | • Secure authentication via identify verification<br>• Does not allow to trace voters from votes<br>• Transparent | • Identity verification service |
| [6] | • Multi-tiered<br>• Two Factor Authentication<br>• Encryption based on public-private keys | • User authentication<br>• Monitoring and auditing for data integrity<br>• Risk of voter to forget their ID, password | • Randomly generated password to use on polling station |
| [8] | • Zcash tokens<br>• Authentication with Challenge-Handshake Authentication Protocol | • Anonymity<br>• Privacy<br>• Transparency | • Challenge-Handshake Authentication Protocol (CHAP)<br>• The voters' email addresses |
| [11] | • Smart contract on Ethereum | • Anonymity | • Asymmetric authentication |

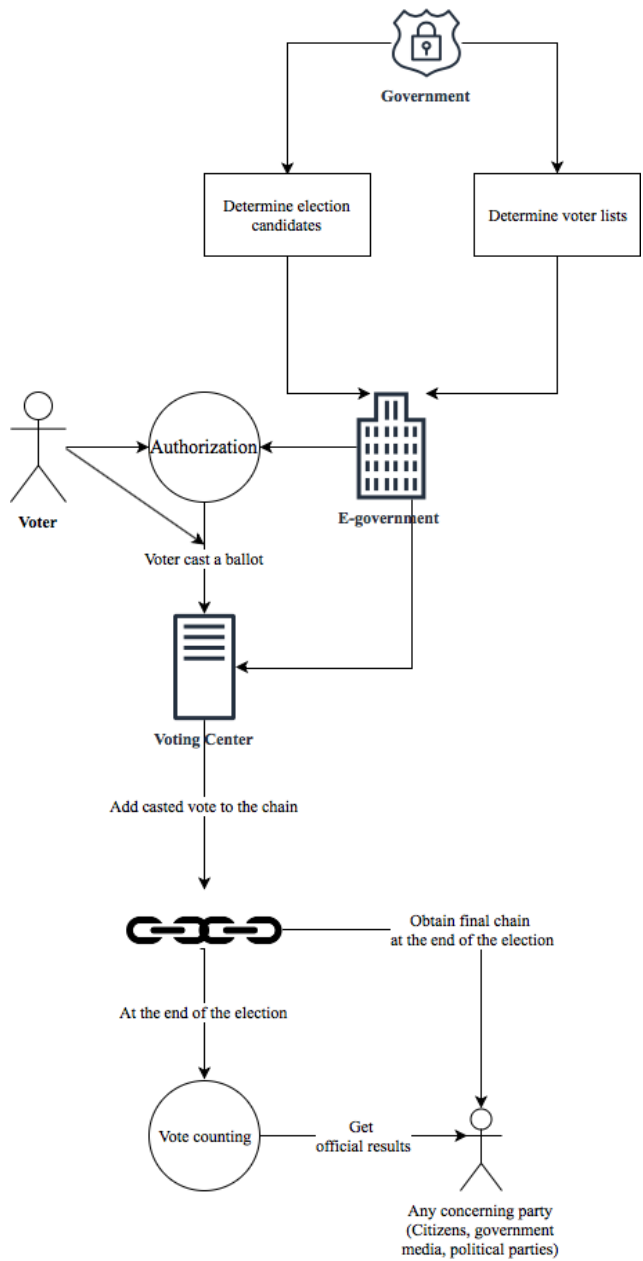

Figure 1: Use case of the blockchain e-voting system

We propose a system that has a leveled structure. There will be different number of levels in that system according to necessities of the country. In order to provide a fast, consistent and secure system, system is designed in a leveled architecture. This number will change from country to country according to features of the country. Reasons behind using a leveled architecture are explained below in detail. Furthermore, consensus of the system is satisfied using DPoS algorithms [12].

If the whole country would have been represented with a single blockchain, synchronization of the system would have a performance issue due to abundance number of ballots and the distance between voting centers. Distance in connected systems is always cause to latency. For a system that includes all the country under the same blockchain, latency between two voting center would be a big problem, because for Turkey, expected latency would be around 100 ms at least. This is a huge value for a system that consist of ten thousand of centers and there would be voting at each center simultaneously. In this case, synchronization of the system would take lots of time. So, in order to decrease latency, chains are distributed over levels. From lowest level to highest level, there will be different chains at each level, and connections between levels will be provided with a secure system.

At the lowest level, there will be a chain that consist of nodes (machines / voting centers) where citizens will perform their voting about election. Due to the relatively less number of nodes in the system, synchronization will take affordable amount of time at the lowest level. When the number of nodes is arranged in a good pattern (i.e. there will not be overload at the chains that will cause enormous latency), system will perform well. Citizen will go to the center and will enter the system with the identity that is provided by the government. We considered to build a system that is working on government's system that will hold the data about citizen for specified voting. If a citizen has not voted yet, citizen will be able to vote one of the candidates. Candidates' will be hold in a database that will also be stored at government related system, because they are already hold. When the authentication process is satisfied, citizen will vote with choosing one of the proposed candidates or blank vote for those who do not want to vote one of the candidates. In this system, proposed candidates will be taken from database that includes relation between ballot boxes and candidates. Thus, there will be only appropriate candidates. In our case, this government is Government of Turkish Republic. We will be using e-devlet system [13]. Since most of the authentication system used across the government related systems are managed by e-government system, it will not be hard to implement this system relying on this system. When the user pass the authentication phase, citizen will see whether he has voted previously or not. If citizen has not voted yet, citizen will choose desired candidate according to the steps explained above.

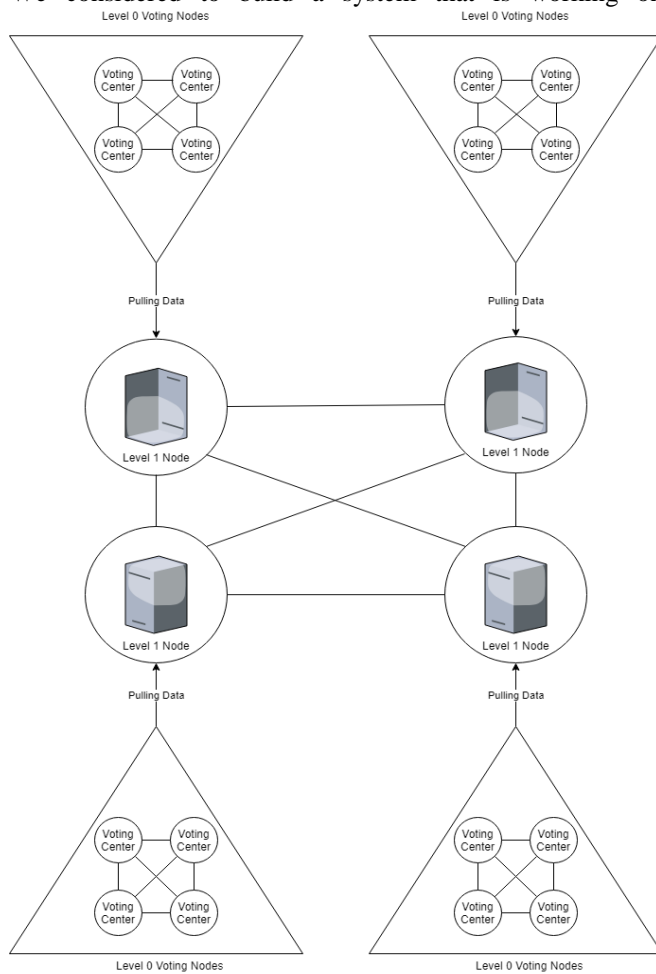

Figure 2: General system architecture

At the second lowest level, there will be a cluster of chains that stores data that are coming from below level. In this level, facilities of blockchain technology are used to make system consistent. For Turkey, we considered that 2 levels will be enough. The system at the second level can have about 700 nodes considering population of the country. That brings a huge performance improvement to the system because the number of connected nodes decrease in this structure in a considerable amount. Additionally, if the node numbers at the 2nd or upper levels are increased, performance increases exponentially. For a country, which has more citizens, level number can be increased in order to decrease collisions between transactions. Consequently, system can be considered as a scalable system.

Communication between levels are ensured using communication protocols. This communication is need to be done periodically. So, there will be a time delay between synchronization of levels. Because, if each vote was considered instantly, there would be a huge bottleneck. This synchronization will provide consistency through the system. For Turkey, according to our calculations, this synchronization time should be 5 minutes. That means, at the end of each 5 minutes period, each node cluster will send the chain data to the upper level node. At this level, data will be synchronized between nodes using a different synchronization algorithm. For this level, we designed an algorithm explained below. You can see the visualization for this two-leveled example in Fig. 2. As you can see there are voting centers which are using same blockchain in their selected area. Also, you can think the voting centers as numerous voting machines but for the sake of simplicity we represent them as voting centers. Moreover, you can see that level 1 nodes are using the same blockchain among the level 1 nodes.

It is stated that vote centers are nodes of blockchains. There will be a file at each node (voting center) that stores the number of data that indicates the number of votes accepted from upper level at previous synchronization step. At every specified time intervals, voting will be stopped for a very short time period (it is expected to be 1 minute for Turkey) in order to synchronize blockchain data between levels. When the data is arrived to upper level from lower chain nodes, it will be checked in order to satisfy consistency. If the consistency of the data is ensured, answer will be a flag that indicates the data is accepted. At this point, nodes at the level 0 will be waiting an answer from level 1 (and level 1 from level 2, so on). If arrived flag tells the votes are accepted, the files at each node (voting centers) will be updated. So, nodes at the lowest level of one of two levels will always know that how many votes have been accepted at above level. Additionally, data will be added to blockchain at the level that the data is arrived (if the communication is between level 0 level 1, indicated level here is level 1). This data will be considered as a transaction block, that means all the new votes (votes coming after from the previously added votes) are considered as a vote cluster and considered as array in computer scientific terms. This vote cluster will be a block that will be added to chain.

At the synchronization phase, if the data coming to the upper level from different machines are inconsistent, that will be a case that should be considered with a care. In this case, if the consensus could not be satisfied, the data will not be accepted from highest level of the two levels and the "decline" flag will be sent to the lower one. In this case, same data should be sent to the upper level again. Until the consistency is satisfied, this procedure will be continued, and with this procedure, consistency will be satisfied at each level.

It is stated above that the all nodes at the lower levels know that the data is accepted if the answer states, so, they can continue working. But in order to satisfy consistency through the system, delays between synchronizations should be arranged very carefully. If the delay between levels becomes a small amount of time, the time spent for synchronization may grow much. On the contrary, if the delay becomes a big amount of time, the data that will be sent between levels takes an enormous size and to transfer this data becomes a problem. So, in order to not bottleneck the system, this delay between levels should be chosen carefully. With the well-designed synchronization times between levels, a high performance providing consistent system would be obtained.

Block structures in levels can be seen below. The election related data are stored in block as shown below through the system. As seen in Fig 3. and Fig 4. we propose two types of blocks: one is for building blockchain at the lowest level of the architecture, that stores candidate info, ballot box info, a "nonce" field that will be used for hashing and a prev_hash info that will be used when creating block that will be added to the blockchain, and other one is consisting of two fields: one is prev_hash field that will be used in order to build blockchain and other one, "lotb" keyword that indicates list of the blocks. Attribute "nonce" is a state of art in the blockchain technology which provides additional security to each hash. For each chain, you select a pattern such as "hash must start with 4 trailing zeros", while creating hash from the block system look to the hash pattern if it does not fit the pattern then it changes the "nonce" string until hash pattern is valid for the chain. It requires more computational power because it will run hash algorithm multiple times until pattern is matched but it also increase the security since if any malicious party create a hash and try to add that block to the chain, it also has to know the hash pattern. Since chain pattern will be different for each election, third parties cannot predict or create a hash that can be accepted.

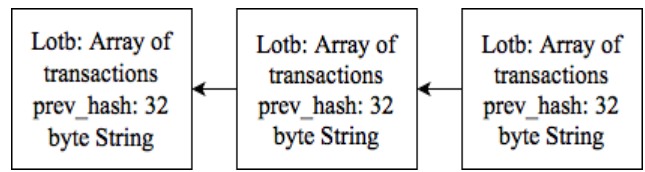

Figure 3: Block structure at level 1 or more

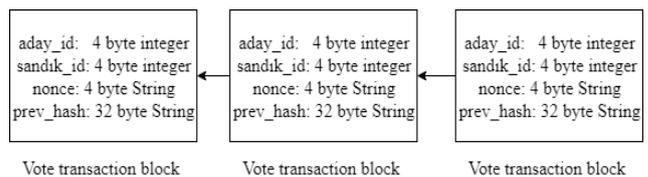

Figure 4: Block structure at level 0 (lowest level)

At the lowest level (level 0), each block will be consisting of one transaction and at each block, whole related information about transaction (in proposed e-voting case, this block indicates a vote) are stored. At the upper levels, the votes coming from one below level are stored in clusters that is sent through different time segments, and all of them are stored as a monolith structure. When this data is arrived, a new block is processed with the prev_hash data and added as a block to the blockchain at described level. When a new block is being tried to add to the chain, according to the Delegated Proof of Stake of Etherium, it is added and data consistency is satisfied. Adding block to the chain is very costly operation. Therefore, there are some implementations and research about this operation. In [14] Ethereum based Smart Contracts system has been discussed for different purposes. Researchers think that smart contracts can be applied in e-voting and this project implements Smart Contracts. Smart Contracts reduce the cost of the transactions and it does not rely on a third party to operate. Turing-completeness feature of the Ethereum allows to create customized and more powerful contracts. More technical details with different inspection methods can be seen in [14].

Summary of the voting procedure can be seen below.

- **Authentication**:
  - Node gets *credentials ← (voter_id, password)* and sends them with its *node_id* to e-government system.
  - E-government system validates user credentials, *validate (credentials, node_id, usersInfo)*.
- **Voting**:
  - Let vote be *v ← vote (voter_id, candidate)*
  - v is added to blockchain, *add (v, chain)*
  - chain information updated for all the voting machines
  - voter's related field is changed to voted in e-government system, *vote (voter_id, userList, true)*.
- **Counting**:
  - Candidates are received from government, *candidates ← getCandidates(candidateList)*.
  - Using highest level chain, votes are counted and winner of the regions are determined, *results ← count (chain, candidates)*.
  - Final blockchain can be distributed to any third party to inspect the anonym votes with.

## IV. ANALYSIS

With the using proposed e-voting system, there will be a system that can be modified for each election easily. Because of system is designed considering candidates at the main voting mechanism and data can be provided however it is wanted. Thus, this system is easily generalizable.

In this system, whole voting information are hold at the highest leveled blockchain so, voting information of the whole country can be reachable instantly at any time after it is synchronized. So, it will not be a problem to explain results of the election, and lots of time that is being spent in order to count votes will be saved. This will change the old and ineffective system and bring a modern and effective system to Turkey, and also lead to save lots of energy and money.

According to calculated statistics considering the data taken from official statistics site of Turkey [15], it is known that there were about 56 million voters at the 2018 general elections in Turkey. Considering this number, it is assumed that 700 clusters (clusters include voting centers under them) would be enough. This is not a random number; it is chosen according to population of the country and bottleneck probabilities at the levels. The geography is a big factor here, due to the latency.

As all the systems, there is a potential threat for the proposed election system. As Yli-Huumo et al. states, if any attacker can get 51% of the computation power of the whole network then it can manipulate the data and this is called 51% attack [16]. In proposed system, node can only be gathered via hacking or stealing the voting machines. Yet, stealing 51% of the voting machines of the country is nearly impossible because in the elections law enforcements are protecting the voting centers therefore any physical theft can be prevented.

Apart from the previously mentioned risks, in case of a disaster citizen who had to vote on the voting machine that has damaged or unavailable to server can be distributed to the nearest voting centers with legal inspections of the supreme committee of elections. Citizens who vote on the voting center that has any problem due to technical problems and disasters can cast their votes in different voting center. Although disaster can damage to the voting machine, blockchain keeps secure all the casted votes. Therefore even in extreme cases during the election day, elections can be completed safely without any doubt.

## V. CONCLUSION AND FUTURE WORK

Democracies depend on trusted elections and citizens should trust the election system for a strong democracy. However traditional paper based elections do not provide trustworthiness. In this paper, we proposed a blockchain based e-voting system which provides trusted, secure and fast voting system for Turkey. Proposed system is suitable to apply in another country whereas integration is a hard work since each country has different laws and election system changes between countries. For the future work, system can be applied for a use case and measurements can be taken to compare if the calculations hold. Synchronization and consensus algorithms can be discussed and improved for better performance and security.

## REFERENCES

[1] R. Hanifatunnisa and B. Rahardjo, "Blockchain based e-voting recording system design," *2017 11th International Conference on Telecommunication Systems Services and Applications (TSSA)*, Lombok, 2017, pp. 1-6.

[2] R. Krimmer, A. Ehringfeld, and M. Traxl, "The Use of E-Voting in the Austrian Federation of Students Elections 2009," Internet: https://pdfs.semanticscholar.org/6b8f/34a5bd3e7eabc7e3a9a3f008187e4415e26a.pdf [Nov. 26, 2018]

[3] "The Geneva Internet Voting System," Internet: https://www.coe.int/t/dgap/goodgovernance/Activities/E-voting/EVoting_Documentation/passport_evoting2010.pdf [Nov. 25, 2018]

[4] F. Þ. Hjálmarsson, G. K. Hreiðarsson, M. Hamdaqa and G. Hjálmtýsson, "Blockchain-Based E-Voting System," *2018 IEEE 11th International Conference on Cloud Computing (CLOUD)*, San Francisco, CA, 2018, pp. 983-986.

[5] S. Ølnes, J. Ubacht and M. Janssen, "Blockchain in government: Benefits and implications of distributed ledger technology for information sharing", *Government Information Quarterly*, vol. 34, no. 3, pp. 355-364, 2017.

[6] A. Barnes, C. Brake, and T. Perry, "Digital Voting with the use of Blockchain Technology," Available: https://www.economist.com/sites/default/files/plymouth.pdf [Nov. 20, 2018]

[7] A. Barnes, C. Brake, and T. Perry, "Digital Voting with the use of Blockchain Technology," Available: https://www.economist.com/sites/default/files/plymouth.pdf [Nov. 20, 2018]

[8] M. Pawlak, A. Poniszewska-Marańda and N. Kryvinska, "Towards the intelligent agents for blockchain e-voting system," *Procedia Computer Science*, vol. 141, pp. 239-246, 2018.

[9] P. Tarasov and H. Tewari, "The Future of E-voting," *IADIS International Journal on Computer Science and Information Systems*, vol. 12, no. 2, pp. 148-165.

[10] Bartolucci, S., Bernat, P., & Joseph, D. (2018). SHARVOT: *Secret SHARe-Based VOTing on the Blockchain. 2018 IEEE/ACM 1st International Workshop on Emerging Trends in Software Engineering for Blockchain (WETSEB)*, 30-34.

[11] A. Reyna, C. Martín, J. Chen, E. Soler and M. Díaz, "On blockchain and its integration with IoT. Challenges and opportunities", *Future Generation Computer Systems*, vol. 88, pp. 173-190, 2018.

[12] B. Wang, J. Sun, Y. He, D. Pang and N. Lu, "Large-scale Election Based On Blockchain", *Procedia Computer Science*, vol. 129, pp. 234-237, 2018.

[13] Internet: https://lisk.io/academy/blockchain-basics/how-does-blockchain-work/delegated-proof-of-stake [Nov. 25, 2018]

[14] Internet: https://www.turkiye.gov.tr/ [Nov. 25, 2018]

[15] Alharby, Maher, and Aad van Moorsel. "Blockchain Based Smart Contracts : A Systematic Mapping Study." *Computer Science & Information Technology (CS & IT)*, 2017

[16] Internet: https://sonuc.ysk.gov.tr/ [Nov. 16, 2018]

[17] J. Yli-Huumo, D. Ko, S. Choi, S. Park and K. Smolander, "Where Is Current Research on Blockchain Technology?—A Systematic Review," *PLOS ONE*, vol. 11, no. 10, 2016.